\documentclass[runningheads]{llncs}
\usepackage{graphicx}
\usepackage{hyperref}
\usepackage{float}
\makeatletter
\newcommand{\printfnsymbol}[1]{%
  \textsuperscript{\@fnsymbol{#1}}%
}    
\makeatother
\begin{document}
\title{Electroencephalogram Signal Processing with Independent Component Analysis and Cognitive Stress Classification using Convolutional Neural Networks}
\titlerunning{EEG Classification using CNN}
%


\author{Venkatakrishnan Sutharsan  \inst{1} \thanks{Equal Contribution} \orcidID{0000-0002-8074-3298} \and \newline
Alagappan Swaminathan\inst{1} \printfnsymbol{1} \orcidID{0000-0002-1589-687X}  \and \newline
Saisrinivasan Ramachandran\inst{1} \orcidID{0000-0002-2905-768X} \and \newline
Madan Kumar Lakshmanan\inst{2} \and \newline
Balaji Mahadevan\inst{1}
}
\authorrunning{Venkat et al.}
%
\institute{Dept. of Electrical and Electronics Engineering, SSN College of Engineering, India \email  {\{venkatakrishnansutharsan16119@eee.,alagappan16011@eee.,saisrinivasan16083@eee.,balajim@\}ssn.edu.in} \newline \and CSIR - Central Electronics Engineering Research Institute, Pilani, Rajasthan, India \newline \email {mklakshmanan@ceeri.res.in}}
%
\maketitle 
\begin{abstract}
Electroencephalogram (EEG) is the recording which is the result due to the activity of bio-electrical signals that is acquired from electrodes placed on the scalp. In Electroencephalogram signal(EEG) recordings, the signals obtained are contaminated predominantly by the Electrooculogram(EOG) signal. Since this artifact has higher magnitude compared to EEG signals, these noise signals have to be removed in order to have a better understanding regarding the functioning of a human brain for applications such as medical diagnosis. This paper proposes an idea of using Independent Component Analysis(ICA) along with cross-correlation to de-noise EEG signal. This is done by selecting the component based on the cross-correlation coefficient with a threshold value and reducing its effect instead of zeroing it out completely, thus reducing the information loss. The results of the recorded data show that this algorithm can eliminate the EOG signal artifact with little loss in EEG data. The denoising is verified by an increase in SNR value and the decrease in cross-correlation coefficient value. The denoised signals are used to train an Artificial Neural Network(ANN) which would examine the features of the input EEG signal and predict the stress levels of the individual.

\end{abstract}
\section{Introduction}
An Electroencephalogram (EEG) signal is that the recording of spontaneous electric pursuit of the mind recorded over a time period. The neurons of the human brain function work through ever-converting the movement of electrical currents across their membranes. These ever-converting currents generate electric and magnetic fields so as to record by using electrodes from the surface of the scalp. The differential potential of the electrodes are then amplified and recorded as the electroencephalogram signal.

The recorded electroencephalogram signal is usually infected through spurious indicators from one-of-a-kind undesirable sources. This form of infection in scientific nomenclature are called as artifact. It's a crucial challenge to get rid of those artifacts from the electroencephalogram (EEG) sources for extra evaluation of EEG. The artifacts are hard to get rid of due to:
i) higher amplitude than the electroencephalogram signal,
ii) the huge frequency tiers of the elements and
iii) because of their variable geographical distribution.

amongst those artifacts, the vast majority of them are Electrooculogram(EOG) signals, which are because of the attention blinking or movement. Eye motion produces electrical interest(EOG signal) this is sufficient vast to be visible inside the electroencephalogram recording. The EOG is a sign that displays the charges distribution among the cornea and the retina which changes during eye motion. Since there is a overlapping of those artifacts over the specified interest data, there may be a massive loss of background electroencephalogram recording. An optimized path of action to accurately rectify this for an electroencephalogram mixed with eye movement is to first discover the EOG pattern and then to smooth the corresponding electroencephalogram signal component rather than cleansing the whole EEG recording. Usually, the reference electrodes are placed over the mastoid bone (which is that the bone in the back of the ear) of each the ears. Merits of this method is that it is reasonably-priced and cost-efficient however suffers from the intense disadvantage of noise from environment. For this reason, this methodology is most popular for low risk functions which includes BMI and many others.

\begin{figure}
\includegraphics[width=\textwidth]{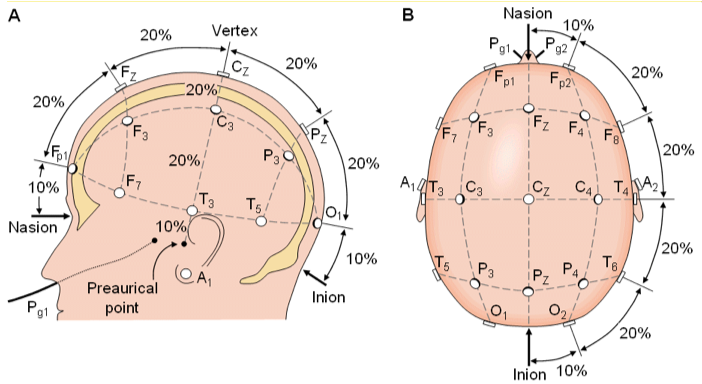}
\caption{10-20 international system electrode placement structure A) lateral view, B) Frontal view.} \label{fig1}
\end{figure}

The International 10-20 electrode system is an internationally identified system to explain and observe the place of scalp electrodes for acquiring electroencephalogram recordings in non-invasive Electroencephalogram. The device relies at the link among the placement of an electrode and the underlying area of the brain, in particular the cortex. The “10” and “20” confer with the truth that the actual distances among adjoining electrodes are either 10\% or 20\% of the entire the front-lower back or right-left distance of the skull. The 10-20 international system of units has the structure as shown in Fig.~\ref{fig1}.

In last few years, a wide range of methods have been carried out for the removal of artifacts in Electroencephalogram. Amongst these artifacts, ocular artifacts are shown to have a huge deterioration of electroencephalogram signal. Many ways to dispose the ocular artifacts had been put forth for a long time, Arjon Turnip et al.(2014) have suggested the elimination of artifacts from electroencephalogram by the use of ICA and PCA. The authors has compared the 2 approaches for eliminating artifacts, i.e. ICA and PCA methodology and from processing with ICA and PCA techniques, It turned that the ICA method is better than PCA in terms of the separation of the electroencephalogram components from blended signal. Anusha Zachariah et al.(2013) have recommended the elimination of artifacts from EEG signal using critical electroencephalogram rhythms. Wavelet decomposition is used inside the pre-processing stage and has proven to increase redundancy and rejection of inappropriate wavelets. This algorithm has 2 step identification of artifact content before and after ICA using kurtosis. Yuan Zou et al.(2012) have planned the elimination of artifacts from electroencephalogram signal using Hierarchical Clustering to split artifacts. S. Jirayucharoensak et al.(2013) have planned the artifact rejection from electroencephalogram using independent component analysis. He delivered a ahead technique to extract useful neural signals with the usage of Lifting Wavelet transform(LWT). Dwi Esti Kusumandari et al.(2015) have counseled the artifacts removal from electroencephalogram using ICA. The authors have as compared two strategies for removing artifacts i.e, JADE and SOBI algorithms. From processing with JADE and SOBI methods, it's far discovered that SOBI technique is higher than the alternative technique in phrases of the separation of the electroencephalogram data from the mixed signal.

There are particular feature for each cognitive stress which are to be used as to classify the EEG signals. Those capabilities/features are Power Spectral Density, wavelet based to name a few. On the grounds that those conventional methods are vulnerable to mistakes, we have used Convolutional Neural Network for classifying EEG Signals. Our Convolutional Neural Network will research the underlying distribution for each signal kind like LAB, COLLEGE, FRIENDS, DOG and CAT. The neural network will do the feature extraction and classification on its own as compared to any other set of algorithms which require each processes to be accomplished one after the other, thus increasing computational speed and reducing the complexity.

\section{Data Acquisition}
\subsection{Brief Layout of The Hardware used for EEG Recording}
Emotiv Epoc Plus EEG Brainwave Headset, which is used to acquire the EEG signals from the scalp. A PC is used to coordinate data flow and also collect data from the sensors attached to it(hence acting like a data acquisition system). A Monitor to view the waveform of the EEG and EOG signal to ensure proper recording of signals. A Holter is a device commonly used to record ECG signals. In this project, it has been used to acquire EOG signals. A pulse generation system which produces impulses of high magnitude and then injects them into the body of the subject. (this is used to synchronize the EEG and EOG signals). The Emotiv Epoc Plus EEG Headset is placed on the head of the person. This headset is connected to the computer by Bluetooth. The EOG electrodes are connected to the Holter through a HDMI cable. Pulse from a programmable pulse generator is given into the electrode. The purpose of giving the pulse is to create markings in the EEG signal waveform corresponding to each window of data.
\begin{figure}
\includegraphics[width=\textwidth]{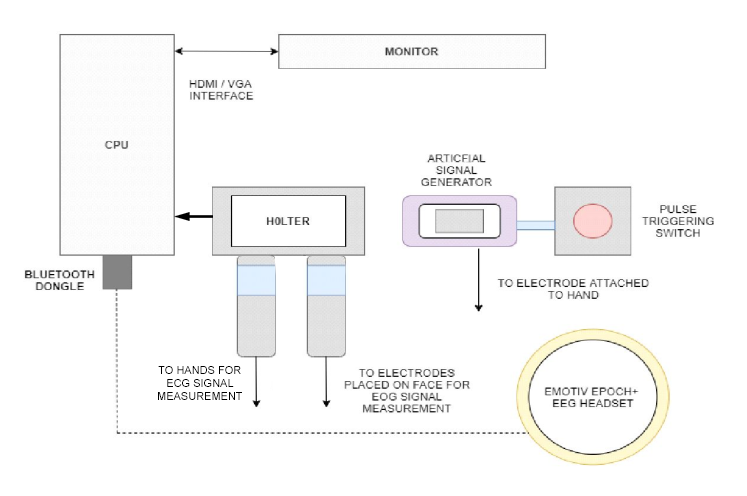}
\caption{Data Acquisition setup} \label{fig2}
\end{figure}

The pulse generated from the external square waveform generator is injected into the ECG signal using two clamps placed on the hands of the subject. The main purpose of injecting a number of pulses into the ECG waveform is to mark the start time of recording and the end time of recording. This is also used for the time synchronizing of the EEG and EOG signals. Since EOG and ECG signals are from the same device, synchronizing ECG signal with the EEG signal inherently synchronizes EEG and EOG signal. 

\begin{figure}
\includegraphics[width=\textwidth]{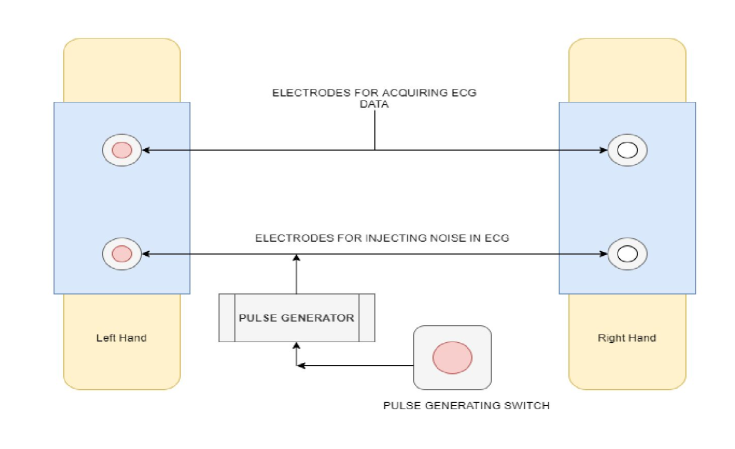}
\caption{Pulse injection setup} \label{fig3}
\end{figure}
\subsection{Procedure for recording EEG signals}
The Emotiv Epoc Plus EEG Headset is placed on the head of the subject. The reference electrodes are placed over the mastoid bone (the bone behind the ear). In the Emotiv Pro software, the connectivity between the headset and the PC is checked. The headset is adjusted to get maximum connectivity. Now, the EOG electrodes are connected to the Holter. The recording is started for both the devices. After 2-3 minutes of recording both EEG, ECG and EOG data, a train of pulses is injected for a duration of 30 seconds – 1 minute. This indicates the starting of EEG signal for processing. The patient is asked to perform certain activity based on the objective of the experiment (like word thinking). Data is recorded for a minimum of 10 minutes and a maximum of 1 hour. Once sufficient data is recorded, a train of pulses with approximately the same duration is given. This serves as a indication for the end time of taking readings from both holter and headset. After the recording is terminated, data from the holter is downloaded into local repository. Headset is removed from the subject's head and the silica gel pads are removed.
\subsection{Pulse waveform generation}
The pulse waveform is generated in an endless loop. To indicate the start of a recording frame, a single pulse of width 2 seconds is given. Then the output is LOW for a period of 6 seconds. Then two pulses are generated with a 50\% duty cycle and time period of 2 seconds.The pulse shown above is generated using an ATMEGA Microcontroller.

\begin{figure}
\includegraphics[width=\textwidth]{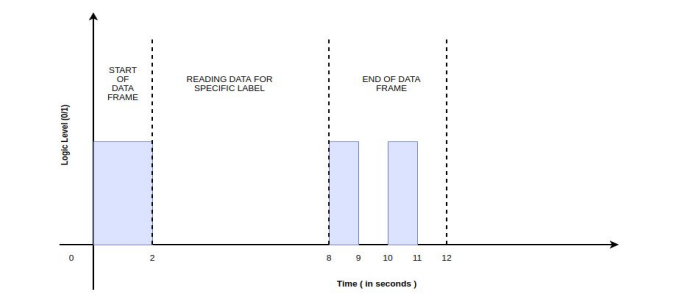}
\caption{Pulse waveform generation setup} \label{fig4}
\end{figure}
\subsection{Collection of data for feature classification}
The subject under observation is asked to look on a monitor. Once a word is displayed on the screen, the subject should think about that in his mind for a certain period of time for 6 seconds. Then the patient is given time to relax his mind and then process repeats till end time of experiment.

\subsection{Procedure for classification of cognitive stress}
The main aim of the experiment is to predict what the person is thinking of. For that, we have to collect EEG data pertaining to some specific objects. Once all the equipment have been arranged, the PULSE\_START\_SWITCH of the interfacing circuit is pressed which starts the pulse generation from the microcontroller. At the same time, a program which would display the commands/instructions is executed. The program is as follows:

The word START appears on the screen, indicating the subject that the experiment is about to begin. This screen is displayed for 2 seconds. Then a label is displayed for a period of 6 seconds. Then the screen displays RELAX for 4 seconds and the loop starts over again.

\section{Analysis Paradigms}
As shown in the (Fig.~\ref{fig5}) there are six steps in this de-noising algorithm. Firstly, the EEG data and the EOG data recorded are pre-processed as per the pre-processing algorithm. After pre-processing the original signals, the second step is to decompose the EEG Signal into multiple Intermediate Components (ICs) using a Source Separation technique called ICA. Thirdly, the correlation coefficient between all the EEG channels and the time synchronized EOG signal are calculated. Then in the forth step, a threshold is set and all the components having correlation value above the threshold are selected. In the fifth step the selected components are scaled down rather than setting it to zero so that the information is not lost completely. Finally, after the scaling down of the components selected as artifact sources the components are projected back into the a group of noise free components. Subsequently, electro-oculogram free signals is reconstructed with Inverse ICA for artifact free components. This noise-free signal is then sliced (640 samples each) into data and its corresponding labels (word associated with the data). This data-label pair is then used to train the Artificial Net to predict each word that the subject is thinking. According to the block diagram this denoising algorithm can be described in detail as follows:

\begin{figure}[hbt!]
\includegraphics[width=\textwidth]{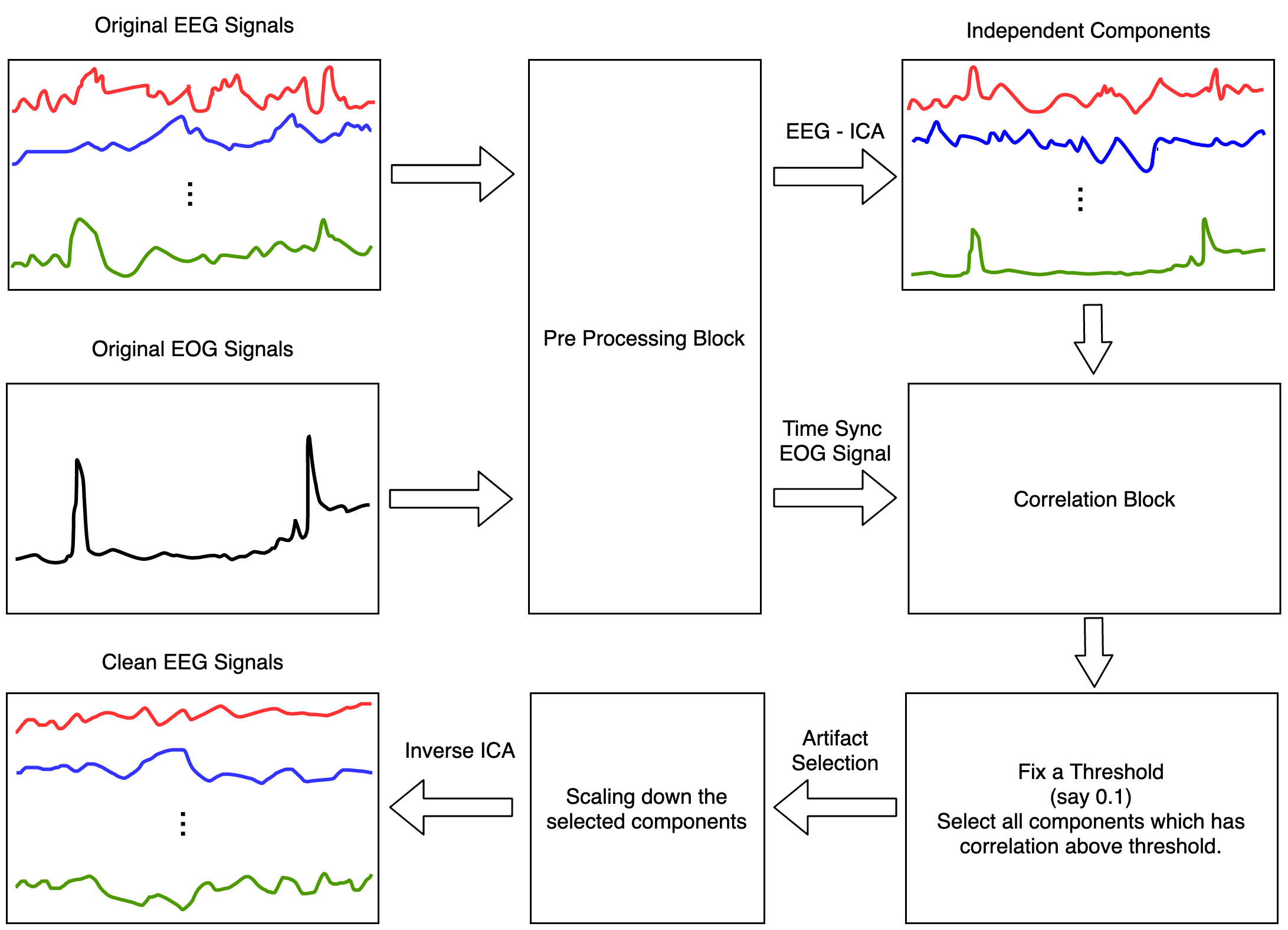}
\caption{Algorithm Process Diagram} \label{fig5}
\end{figure}

\subsection{ Preprocessing}
\begin{figure}[h]
\includegraphics[width=\textwidth]{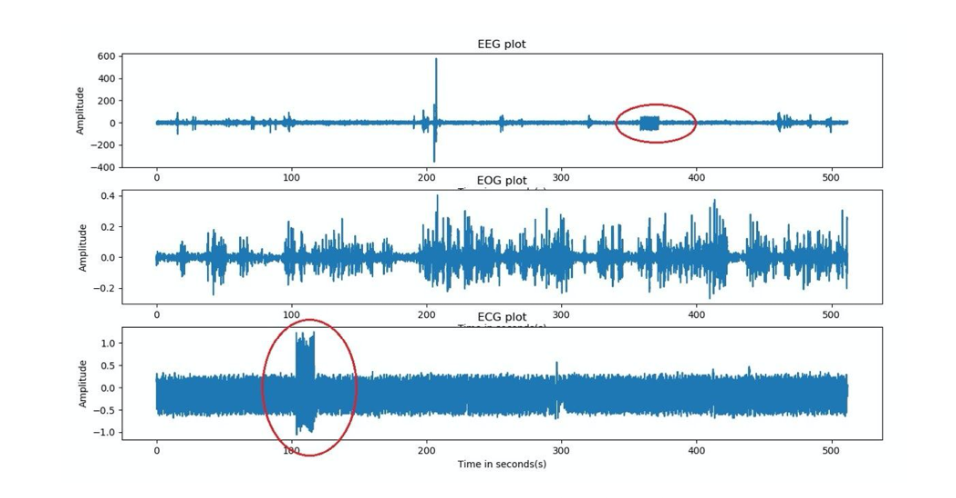}
\caption{Plot of the total recorded signals.} \label{fig6}
\end{figure}

The preprocessing step consists of two main process which include Time Synchronization and Band-pass filtering. Devices which are used to capture the signals of electroencephalogram and electro-oculogram have different starting time, To synchronise the time frame between both the device's signal, we use artificially generated pulses.  A plot of the recorded signal is shown in Fig.~\ref{fig6}.

\begin{figure}[hbt!]
\includegraphics[width=\textwidth]{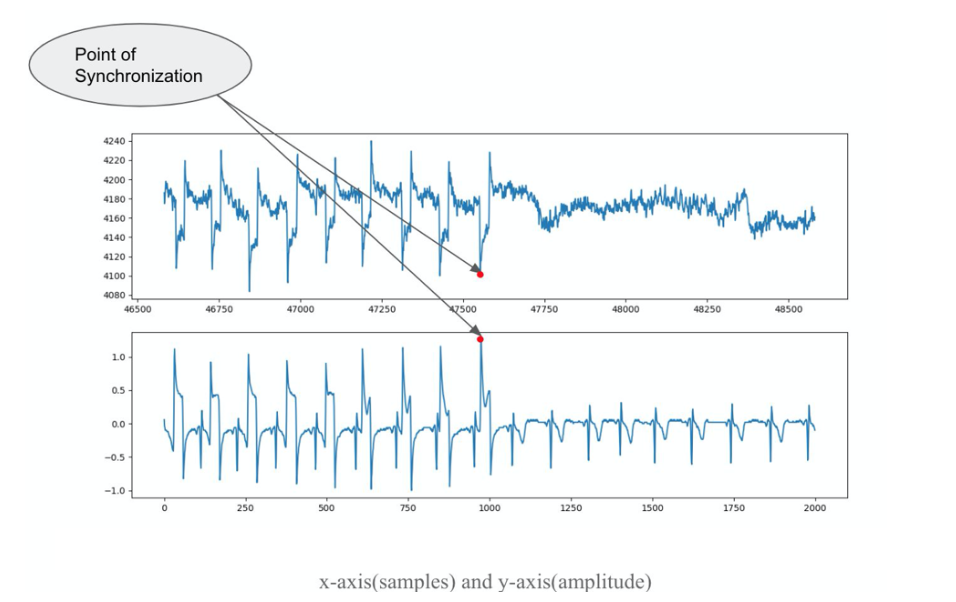}
\caption{Point of synchronization.} \label{fig7}
\end{figure}

\begin{figure}[h!]
\includegraphics[width=\textwidth]{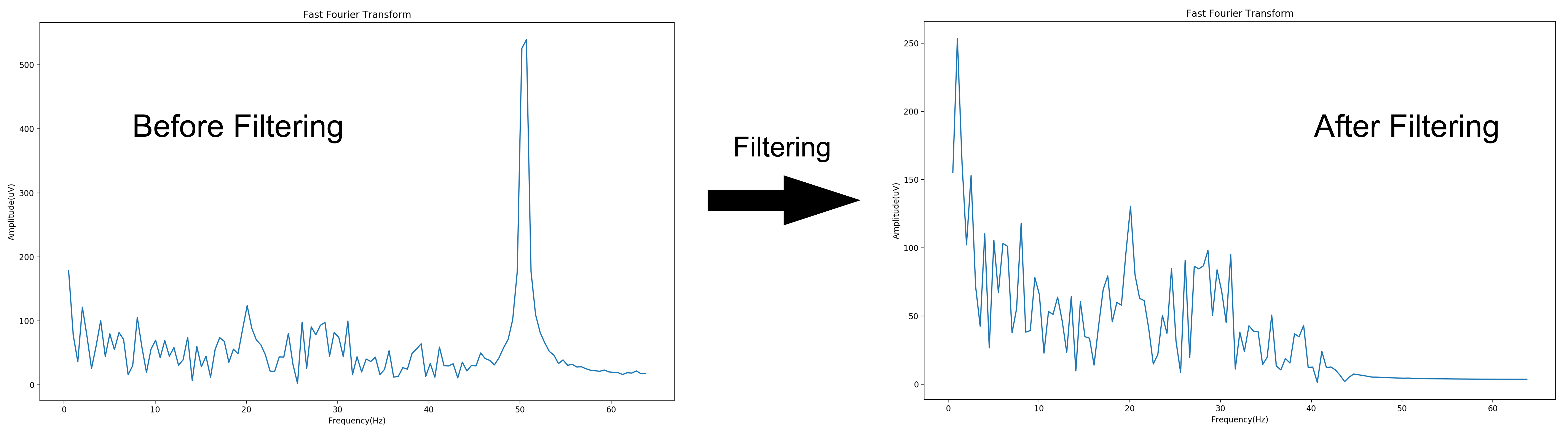}
\caption{Butterworth filter Plot} \label{fig8}
\end{figure}

Pulses used here are artificial impulse signals inoculated into both the EEG and EOG signals during the start and stop of the process. Since the two devices cannot be started together and therefore a lag between 2 devices can cause inaccurate noise detection. Butterworth Filter is used for preprocessing to remove artifacts present above and below region of interest. It is a signal filtering mechanism designed to have a frequency reaction as even as possible in the the bandwidth of the pass band region. This filter is also known as the maximally flat magnitude filter. It was defined originally by physicist Stephen Butterworth. The process is done for a sliding window of 20 sec. The point of synchronization after the peaks are synchronized are shown in Fig.~\ref{fig7}. A plot of the Fourier Transform before and after pre-processing is illustrated in Fig.~\ref{fig8}.

Program Algorithm: Data from the devices are imported into the program. Plot the whole recording for both the devices. Input the indexes of the synchronizing point of both the signals. Slice/crop the signals according to the indexes provided by the user. Filter the new data(Time Synchronized) using butter-worth filter (5th order - bandpass(0.1Hz to 40Hz) .Write the time synchronized and filtered data into a CSV file.

\subsection{Independent Component Analysis} ICA is a well established technique on decomposing a signal into independent sources using kurtosis as the cost function. Independent Component Analysis is an effective technique to de-mix linear mixtures of signals into underlying independent sub-components. ICA utilises higher order statistics like kurtosis to seek out the independent components. ICA is an extension of a statistical model called Principal Component Analysis (PCA). Main supremacy in ICA is that it extracts the sources by investigating the statistical independence underlying the recorded data .Thus, it involves higher order statistics to recover statistically independent signals from the observations of an unknown linear mixture. There are 3 conditions for the Independent Component Analysis algorithm. First, it should be a linear combination of the originating signals. Second, original signals ought to be statistically independent and at last, the independent components should be Non-Gaussian. For instance, Let A be the mixing matrix, which is a square matrix of order two. Now, the mixed signal vector X is a result of the product between the source signal vector S and the mixing matrix A, given by the equation below:
 \begin{equation}
     X=A*S
    \label{eqn1}
 \end{equation}    
The core objective of ICA is to find the originating/source signals from the mixed signals. This is done by discovering the de-mixing matrix (given by W).
The de-mixing equation is given as:
\begin{equation}
    Y=A^\textit{-1} *X
    \label{eqn2}
\end{equation}

\subsection{Computation of Correlation value}
The correlation value between two functions define the trend between two variables on how strong the pair of variables are related. The correlation used here is Spearman Correlation named after Charles Spearman. The Pearson correlation calculates linear relationships whereas the Spearman correlation determines monotonic relationships. The value usually lies between +1 and -1 where the +1 shows positive trend that is both variables move in same direction and -1 shows negative trend where the 2 variables move in opposite direction while 0 show no relation. The correlation coefficient was found between each channel of EEG signal and EOG data. The absolute estimate of the correlation is utilized in this research study as to avoid one particular limitation of ICA and the change in polarity of the electrodes. The value is calculated using the formula presented below:

\begin{equation}
    \rho = 1 - \frac{6\sum{d_i^2}}{n*(n^2-1)}
    \label{eqn3}
\end{equation}

where d is the difference between the rank of the value in the corresponding dataset.


\subsection{Selection of Artifactual Components} 
The eye artifact is found by estimating the component which closely resembles the EOG data.The correlation values calculated using a statistical Spearman correlation is used to select the artifactual component by thresholding a value. Due to practical problems such as voltage bias, the correlation value can change from device to device. The thresholding value was found by experimentation to be 0.1 . The Independent Components having correlation above the threshold is selected as EOG-related components. Additionally optical examination is doneto supervise the selection of all the components.

\subsection{Reducing of Artifact effect on signal}
The selected Independent Components are through a process of scaling down to reduce the effect on EEG signal. This scaling is done in such a way that the component having the most correlation with the EOG data is scaled down more than the component having less correlation. This can be explained by the 2 formulas defined below
\begin{equation}
component = component * (1 - 2*correlation) 
\label{eqn4}
\end{equation}
\begin{equation}
component = component * (1 - correlation)
\label{eqn5}
\end{equation}

The equation ~\ref{eqn4} is used when the correlation is between 0.1 and 0.5 while equation ~\ref{eqn5} is used when correlation is above 0.5, but the latter formula is applied as an exceptional case. This process of scaling the Indpendent Components helps in reducing the noise in the signal while retaining the feature of the EEG signal. Thus, the process of scaling down allows the de-noising algorithm to remove the noise part and not affect the characteristics of the signal.

\subsection{Signal Reconstruction}
The artifact signals are selected as EOG-linked ICs and their effect are reduced using the scaling algorithm. After the EOG-linked ICs are scaled down the set of components are projected back on to the same space (Inverse ICA) as the original components, therefore having EOG-free EEG signal.

\subsection{Signal learning using Convolutional Neural Network}
Apart from the denoising algorithm, in this section we will be training a Convolutional Neural Network to classify the input brain waves. The convolutional layer is the center of a Convolutional Neural Network. The layer's specific parameters comprise of a bunch of teachable channels (or portions), which have a little open field, yet stretch out through the full profundity of the information volume. During the forward pass, each channel is convolved across the batchsize, channel and tests of the info volume, processing the scalar item between the sections of the channel and the information and creating a 1-dimensional actuation guide of that across channels. Therefore, the organization learns channels that actuate once it recognizes some particular assortment of highlight at some spatial situation inside the info. The signs recorded are then parted with regards to the name and their relating signal for instructing. We have utilized 80\% of the recorded information for preparing the neural organization and 20\% to check the organizations exactness. Convolutional Neural Networks are normally utilized for pictures (2 Dimension) while this can be utilized in 1-D space (time-arrangement information) to perceive designs in mind waves. In the Classifier, every one of the thick layers utilize the ReLU (corrected straight unit) as the actuation work aside from the last yield layer(Sigmoid). The reduction of the samples is by 3 stages of convolution (kernel-3, stride-1, padding-0 along with batch normalization and max pooling) with the ultimate layer giving three hundred channels. Dense layers have 23400 → 1024 (in the first hidden layer), 1024 → 512 (in the second hidden layer), 512 → 256 (in the third hidden layer) and 256 → 5 (in the fourth hidden layer) mapping to each of the 5 words.

\textbf{}
\section{Result}
\subsection{Quantitative Validation Measures}
After processing the EEG signals there are standard validation measures such as Signal Noise Ratio and Spearman Correlation. Signal Noise ratio is one standard methodology to demonstrate the signals data over noisy data. The SNR of the signal can be found by 
\begin{equation}
    SNR = 10*log(Pow(S)/Pow(N))
    \label{eqn6}
\end{equation}
where S relates to the Signal and N happens to be the Noise. The power can be calculated like the variance of the time signal by Hjorth Activity. The increase in SNR value show a strong increase in signal which is useful in applications such as Sleep Study for Medical Experimentation.

The Correlation is also another useful parameter to measure the degree to which two variables move in relation to each other. This can be calculate by using the Spearman Correlation using the equation \ref{eqn3}.

For neural network the accuracy of the prediction is found out by dividing the count of images predicted correctly by the total number of images subjected to the network.
\begin{equation}
    Accuracy = (Total\ correctly\ predicted\ images)/(Total\ images\ to\ the\ network)
\end{equation}
\subsection{Experimental Data}
\begin{figure}[!hbt]
\includegraphics[width=\textwidth]{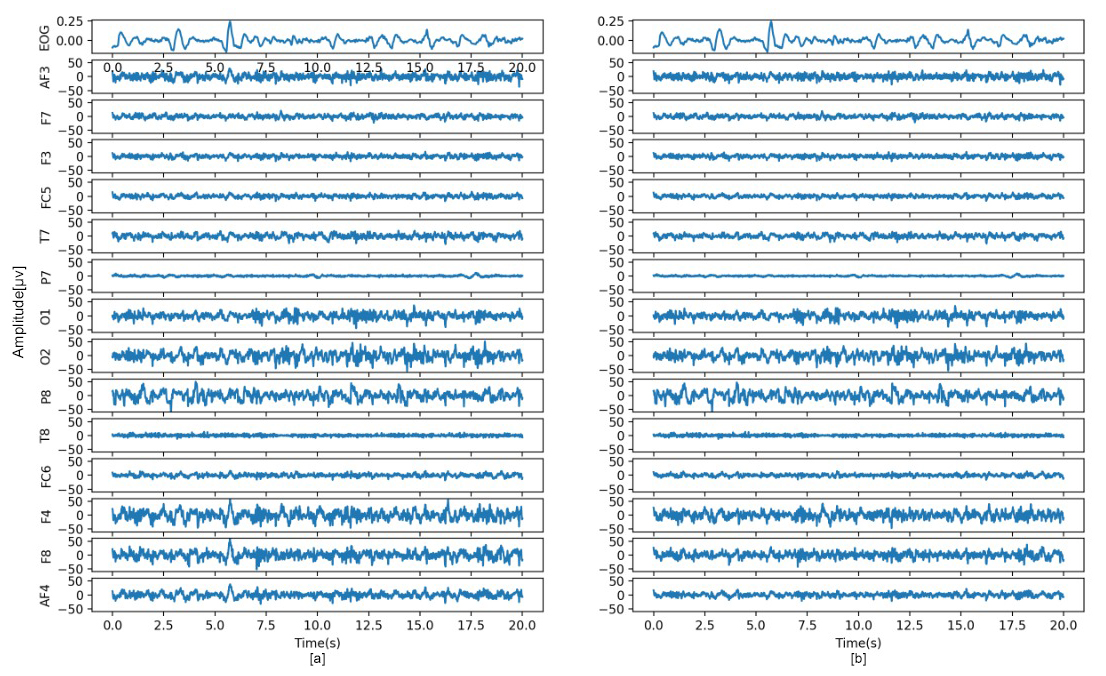}
\caption{(a) A typical example of a 14-channel electroencephalogram signal with effect of eye-blinks and the first channel is EOG reference signal. (b) The 14-channel Artifact free EEG signal with first channel EOG reference signal.} \label{fig9}
\end{figure}

In this study, the planned technique is used to get rid of EOG-related artifacts from electroencephalogram signals. A classic multiple channel of true electroencephalogram signals which are polluted with ocular blinks and with the original noise free electroencephalogram signals once EOG-related signal removal are illustrated in Fig.~\ref{fig9}. Especially, the channels close to the eye has shown massive amplitude of noise signals corresponding to the electro-oculogram. The first signal bandwidth containing channel refers to the EOG signal.

We can use correlation values linking the electroencephalogram channels and electro-oculogram signal to estimate whether the EOG-related signals were removed or not. The Average correlation value of each electrode before and after denoising was calculated for the data and is illustrated in Table ~\ref{tab1}. It can be easily understood that the correlation value after denoising is less than the correlation value before denoising. The difference in correlation before and after denoising is dominant in the channels where the EOG related signals have high influence. The channel have lower influence of eye artifact has less denoising effect, thus preserving the details of the signal.

\begin{table}[!hbt]
\centering 
\caption{Value of Correlation before and after Denoising.}\label{tab1}
\begin{tabular}{|c|c|c|c|c|c|c|c|c|c|c|c|c|c|c|}
\hline
Ch. Name & AF3 & F7 & F3 & FC5 & T7 & P7 & O1 & O2 & P8 & T8 & FC6 & F4 & F8 & AF4 \\
\hline
Before &  0.336 & 0.041 & 0.233 & 0.242 & 0.015 & 0.107 & 0.063 & 0.043 & 0.092 & 0.040 & 0.170 & 0.294 & 0.288 & 0.450 \\
\hline
After &  0.139 & 0.019 & 0.103 & 0.142 & 0.005 & 0.064 & 0.001 & 0.023 & 0.040 & 0.008 & 0.014 & 0.039 & 0.020 & 0.186 \\
\hline
\end{tabular}
\end{table}
Additionally, comparing the waveforms and the correlation value before and after denoising, it can be found that the denoising algorithm has effectively denoised the EEG signal without loss of too much information thus preserving the data for further analysis.

As we don't know the noise and feature signals before denoising we cannot calculate the SNR value before denoising. But after denoising since we know the denoised signal, the noise can be calculated using the Formula ~\ref{eqn7}.
\begin{equation}
    Noise = Input\textit{ }EEG\textit{ }Signal - Clean\textit{ }EEG\textit{ }Signal
    \label{eqn7}
\end{equation}
With the noise and the clean EEG signal, we can calculate the SNR value by using the Formula ~\ref{eqn6}. The average SNR value for all the subjects for the corresponding electrodes are shown in Table ~\ref{tab2}.

\begin{table}
\centering
\caption{SNR Value for all the 14-electrodes after Denoising.}\label{tab2}
\begin{tabular}{|c|c|c|c|c|c|c|c|c|c|c|c|c|c|c|}
\hline
Ch. Name & AF3 & F7 & F3 & FC5 & T7 & P7 & O1 & O2 & P8 & T8 & FC6 & F4 & F8 & AF4 \\
\hline
SNR value &  11.56 & 45.58 & 24.26 & 25.05 & 22.82 & 17.86 & 42.10 & 29.47 & 37.74 & 46.50 & 14.97 & 2.67 & 3.80 & 4.12 \\
\hline
\end{tabular}
\end{table}
The SNR value show that there is a significant increase in signal strength. The SNR is found to be high for electrodes which have a higher decrease in correlation between the corresponding electrode and the EOG signal.

\begin{figure}
\includegraphics[width=\textwidth]{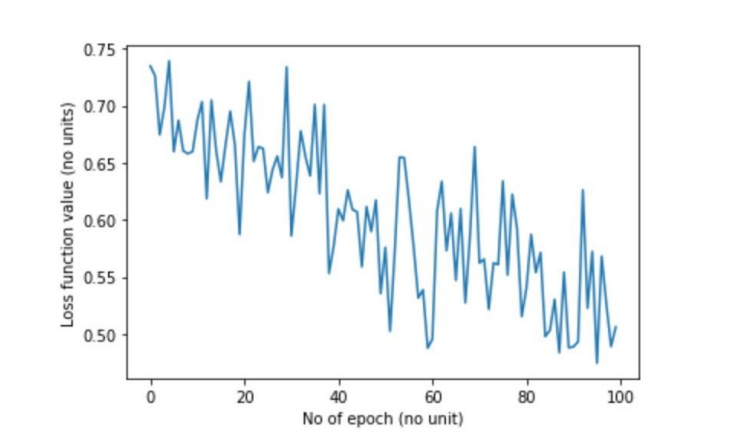}
\caption{Loss curve of the neural network} \label{fig10}
\end{figure}

Finally, we have successfully developed a convolutional neural network which can predict the state of mind (thinking of DOG, CAT, COLLEGE, LAB and FRIENDS). After the neural network was trained, we subjected the test dataset to the network for label prediction and was found to have an accuracy of 89.91\%. This means that the Neural Net is able to precisely forecast one from the list of labels for 90 of every 100 images in the dataset. The learning loss of the neural network is shown in Fig.~\ref{fig10}.

\section{Acknowledgement}
We would like to thank the Council of Scientific and Industrial Research - Central Electronic Engineering Research Institute for supporting us with the lab and for providing us the necessary hardware equipments for data collection. We would also like to thank the 3 Homo sapiens who were in good physical health(Gentlemen, average age 21, spanning 20-22 years old) who willingly took on to active participation in this study.
\section{Conclusion}
In this research paper, a technique proposed for cognitive stress classification was assessed by true and original electroencephalogram signals acquired from Emotiv Epoc Plus headset. The results indicate that the proposed method can eliminate EOG-related signals from acquired data with minimal loss of information and predict the cognitive stress. This can be seen by the increase in SNR value and also by the decrease in cross-correlation coefficient value. The method proves to remove dominant eye related activity as the effect of EOG signal is higher compared to ECG and EMG signals. Also, The results from the Convolutional Neural Network indicate a very high level of accuracy in classifying the cognitive stress.

\section*{Declaration}  




\subsection*{Authors' contributions}
VKS finished the experimental research, participated within the evaluation of the signals, participated in the process of analysing the various signals recorded and drafted the manuscript. AS carried out the experimental studies, participated in the process of analysing the various signals recorded and drafted the manuscript. SR conducted literature survey and performed information acquisition. MKL coordinated the ethics application, came up with the direction for the evaluation of the conclusion, and confirmed the validity of the outcomes. BM coordinated the ethics utility, model recommendations for the evaluation of the terminal outcome, and confirmed the validity of the effects. All authors read and accepted the very last manuscript. All authors read and permitted the very last manuscript.

%
%
%

\end{document}